\documentclass{moriond}

\usepackage{amsmath}
\def\Journal#1#2#3#4{{#1} {\bf #2}, #3 (#4)}

\def\jchem{\em J. Chem. Phys.}
\def\AA{\em A\&A}
\def\AJ{\em AJ}

\def\be{\begin{equation}}
\def\ee{\end{equation}}
\def\bea{\begin{eqnarray}}
\def\eea{\end{eqnarray}}

\newcommand{\Photo}{\includegraphics[height=35mm]{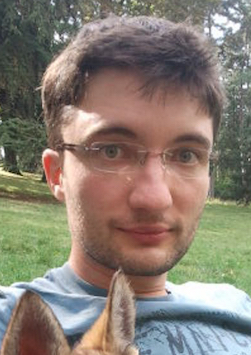}}

\begin{document}
\vspace*{4cm}
\title{Nonideal self-gravity and cosmology: the importance of correlations in
  the dynamics of the large-scale structures of the Universe}

\author{P. Tremblin}

\address{Maison de la Simulation, CEA, CNRS, Univ. Paris-Sud, UVSQ, Universit\'e
  Paris-Saclay, F-91191 Gif-sur-Yvette, France}

\maketitle\abstracts{Inspired  by the  role of  correlations in  the statistical
  mechanics  of nonideal  self-interacting  fluids, we  suggest that  unresolved
  sub-structures (i.e. correlations) have to be taken into account in the Virial
  theorem of  self-gravitating astrophysical systems. We  demonstrate that their
  omission leads to a missing mass problem by using the semi-analytic polytropic
  solutions of  the Lane-Emden  equation.  This problem  suggests to  extend the
  Friedmann equations to the nonideal regime by taking into account correlations
  in the dynamics of the expansion.  The increase of correlations induced by the
  formation  of   the  large-scale   structures  could  explain   naturally  the
  accelerated expansion of the Universe in such a paradigm.}

\section{The role of correlations in statistical mechanics}

The statistical approach  of the classical problem of a  system of N-interacting
particles (with $N\rightarrow \infty$) of mass $m$  in a volume $V$, is based on
the  so-called BBGKY  hierarchy  \cite{yvon,kirkwood}.  In  this hierarchy,  the
probability  function  to   find  one  particle  at  $\vec{r}$   is  defined  by
$P_1(\vec{r})$  while the  probability to  find  one particle  at $\vec{r}$  and
another one at $\vec{r}^\prime$ is given by $P_2(\vec{r},\vec{r}^\prime)$. These
probabilities yield the definition of the radial distribution function as, for a
homogeneous, isotropic fluid:
\begin{equation}
  g(||\vec{r}-\vec{r}^\prime||) = V^2 P_2(\vec{r},\vec{r}^\prime),
  \label{eq:gr}
\end{equation}
or equivalently the correlation function, $\xi(||\vec{r}-\vec{r}^\prime||)$:
\begin{eqnarray}
  \xi(||\vec{r}-\vec{r}^\prime||)    &=&    V^2(P_2(\vec{r},\vec{r}^\prime)    -
  P_1(\vec{r})P_1(\vec{r}^\prime))\cr &=& g(||\vec{r}-\vec{r}^\prime||) -1.
  \label{eq:corr}
\end{eqnarray}
For an ideal fluid, in  the absence of correlation, $P_2(\vec{r},\vec{r}^\prime)
=  P_1(\vec{r})P_1(\vec{r}^\prime)$,  i.e.   $\xi(||\vec{r}-\vec{r}^\prime||)=0$
and  $g(||\vec{r}-\vec{r}^\prime||)=1,\,\forall  (\vec{r}, \vec{r}^\prime)$.  In
contrast, in the presence of correlations, $\xi(||\vec{r}-\vec{r}^\prime||)\ne0$
and $g(||\vec{r}-\vec{r}^\prime||)\ne1$.  The presence  of correlations yields a
non-ideal  contribution impacting  all  thermodynamic quantities  in the  fluid,
including also interaction energies whose definition from the N-body dynamics is
the sum over all distinct pairs of particles:
\begin{equation}
  \langle           H_\mathrm{int,Nbody}\rangle           =           \sum_{i<j}
  \phi(||\vec{r}_i-\vec{r}_j||),
  \label{sum}
\end{equation}
which  corresponds  exactly  to  an  integral  over  the  two-point  probability
distribution function:
\begin{equation}
  \langle    H_\mathrm{int,Nbody}\rangle     \approx    \frac{N^2}{2}\iint_{V,V}
  P_2(\vec{r},\vec{r}^\prime)\phi(||\vec{r}-\vec{r}^\prime||)dV dV^\prime,
  \label{Eint}
\end{equation}
with $\phi(r=||\vec{r}-\vec{r}^\prime||)$ the pair potential.

\section{The missing-mass problem induced by the omission of correlations}
\label{sect:polytrop}

The  Virial theorem  in astrophysics  is usually  applied in  a simplified  way,
ignoring correlations and sub-structures by using only the total mass $M$ inside
a given volume  $V=4\pi R^3/3$ with a gravitational interaction  energy per unit
mass proportional to $-GM/R$.

\begin{figure}
  \begin{centering}
    \includegraphics[width=0.49\linewidth]{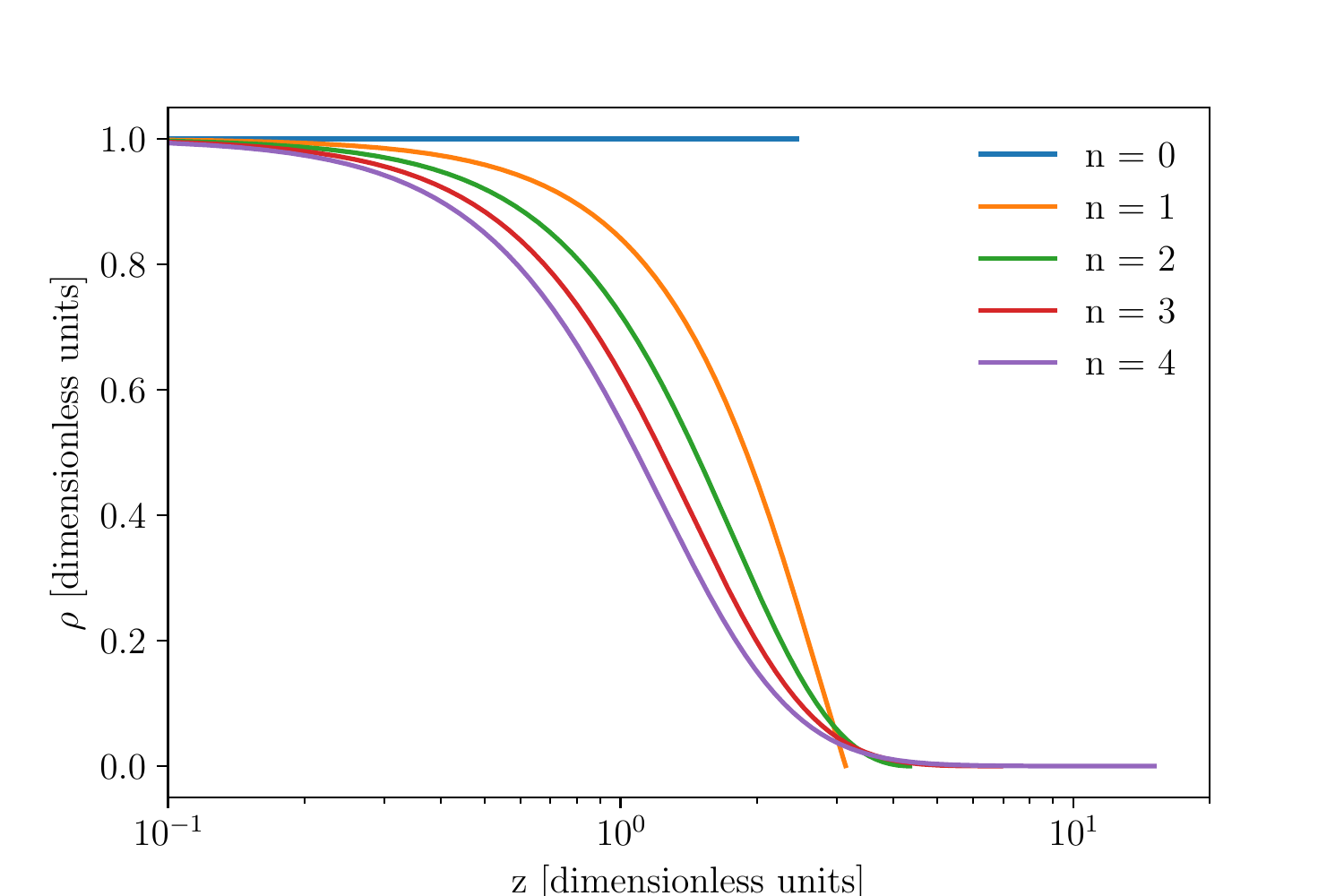}
    \includegraphics[width=0.49\linewidth]{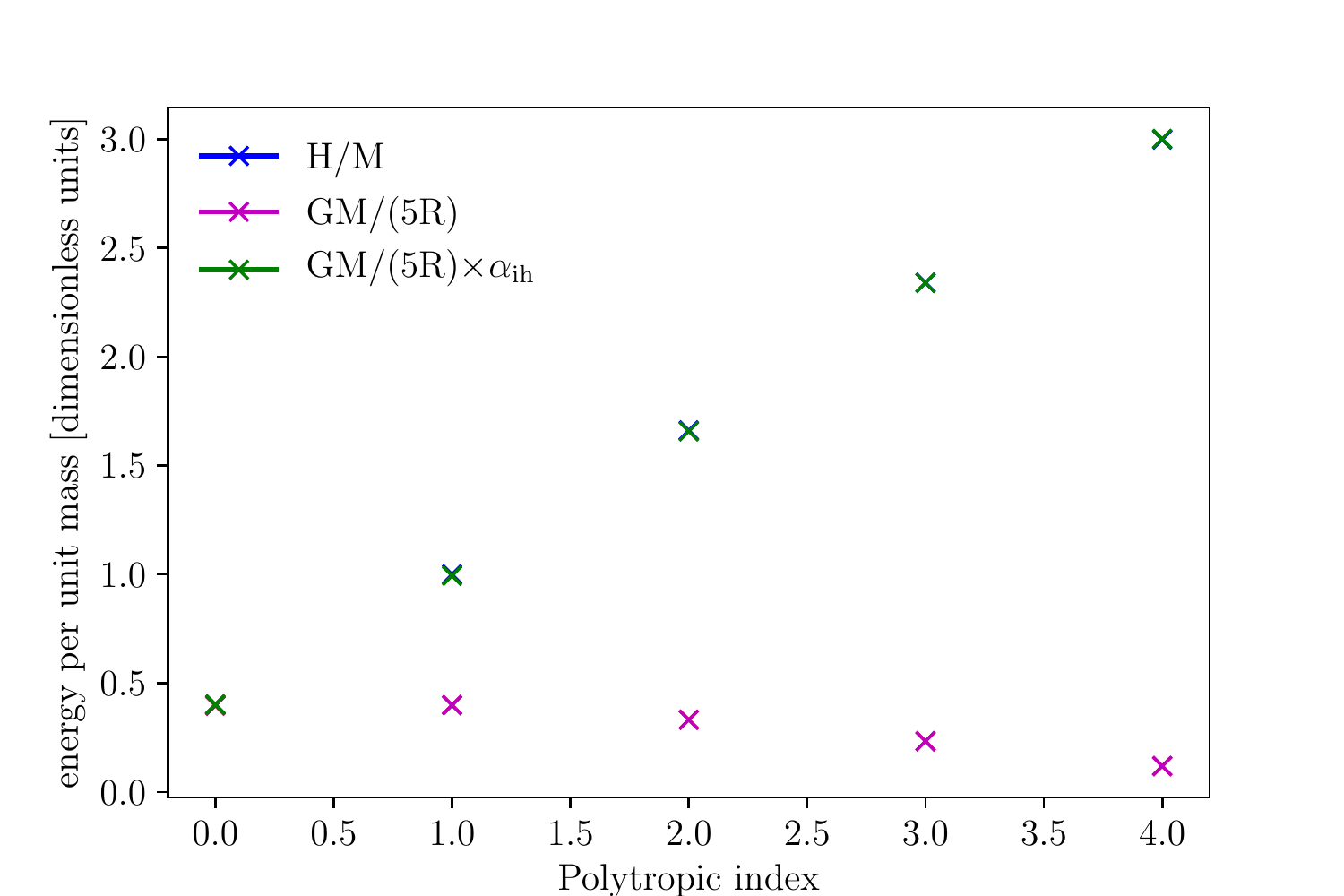}
    \caption{Polytropic  solutions.  Left:  Density profiles  solutions  to  the
      Lane-Emden equation  with different polytropic indexes  $n$.  Right: Total
      enthalpy per  unit mass compared  to gravitational interaction  energy per
      unit mass assuming either a  homogeneous density profile or accounting for
      inhomogeneities.        Blue       and       green       crosses       are
      indistinguishable.} \label{fig:polytrops}
  \end{centering}
\end{figure}

In order  to show the  importance of inhomogeneities  in the virial  theorem, we
apply  it  as a  simple  semianalytical  example to  the  case  of a  polytropic
structure.      In     this     case,      we     can     explicitly     compute
$P_2(\vec{r},\vec{r}^\prime)$  by  semianalytically computing  an  inhomogeneous
density field and compare the result with a homogeneous assumption on the virial
theorem.  A  polytropic stellar  structure can be  calculated by  the Lane-Emden
equation,
\begin{equation}
  \frac{1}{z^2}\frac{d}{dz}\left( z^2 \frac{dw}{dz}\right) + w^n = 0,
\end{equation}
with  $\rho =  \rho_c  w^n$, $P=K\rho^\gamma$,  $n=1/(\gamma-1)$ the  polytropic
index,   $z=Ar$,  and   $A^2=(4\pi  G/(n+1))\rho_c^{1-1/n}/K$.    We  choose   a
dimensionless unit  system in  which $\rho_c=K=G=1$.  By solving  the Lane-Emden
equation for  different polytropic indexes  $n<5$, we can  compute inhomogeneous
density  profiles  that  have  a  finite  radius $R$  (see  the  left  panel  of
Fig.~\ref{fig:polytrops}). For  a polytrope,  the following virial  theorem (per
unit mass) reads
\begin{eqnarray}
  H/M &= 2H_\mathrm{int}/M, \cr H &= \int_V \frac{\gamma}{\gamma-1}P dV,\cr M &=
  \int_V \rho dV,
\end{eqnarray}
where $H$ is the total enthalpy, $M$  is the total mass, and $H_\mathrm{int}$ is
the gravitational interaction  energy with the zero of  potential energy defined
at  the surface  of the  star.  For  a homogeneous  density profile,  with $\rho
=M/(4\pi R^3/3)$, we obtain
\begin{equation}\label{eq:ehomo}
  \frac{H}{M} = \frac{GM}{5R}.
\end{equation}
We now assume that an observer has  only access to the total enthalpy, mass, and
radius  of different  objects  and  does not  know  whether  these objects  have
substructures.   The naive  assumption, as  explained  above, is  to ignore  the
possible  inhomogeneities and  assume  that  the density  is  constant, $\rho  =
M/(4\pi R^3/3)$, hence  using Eq.~\ref{eq:ehomo} for the  virial equation.  This
assumption, however, is only correct for  a polytropic index $n=0$.  As shown in
the right panel of Fig.~\ref{fig:polytrops}, for profiles $n\ne 0$, the error on
the  total energy  is  significant and  can  be up  to a  factor  25 for  $n=4$.
Therefore, the  assumption of a  homogeneous profile within the  structures will
lead to  a missing-mass problem.   For a  polytropic structure, however,  we can
calculate the inhomogeneous density  profile semianalytically and thus calculate
exactly  the   contribution  of  inhomogeneities   to  the  total   energy.   We
characterize  this   contribution  by  an  inhomogeneous   amplification  factor
$\alpha_\mathrm{ih}$ , which can be calculated exactly,
\begin{equation}
  \alpha_\mathrm{ih}  =  \frac{  N^2  \iint_{V,V}  P_2(\vec{r},  \vec{r}^\prime)
    \phi(|\vec{r}-\vec{r}^\prime|)   dV   dV^\prime}{N^2   \iint_{V,V}   1/(4\pi
    R^3/3)^2 \phi(|\vec{r} - \vec{r}^\prime|) dV dV^\prime},
\end{equation}
with  $V=4\pi R^3/3$.   In our  calculation, $N^2  P_2(\vec{r},\vec{r}^\prime) =
N^2P_1(\vec{r})P_1(\vec{r}^\prime) = \rho(\vec{r}) \rho(\vec{r}^\prime)$ for the
inhomogeneous  structure at  the  numerator, while  the homogeneous  calculation
corresponds         to         $N^2        P_2(\vec{r},\vec{r}^\prime)         =
N^2P_1(\vec{r})P_1(\vec{r}^\prime)  = N^2/V^2$,  with $P_1(\vec{r})=1/V$  at the
denominator.    We   can  now   correct   Eq.~\ref{eq:ehomo}   to  account   for
inhomogeneities,
\begin{equation}\label{eq:ehomo2}
  \frac{H}{M} = \frac{GM}{5R} \alpha_\mathrm{ih}.
\end{equation}
As shown  in the bottom panel  of Fig.~\ref{fig:polytrops}, when this  factor is
taken into  account, we  exactly recover  the correct  enthalpy and  the correct
gravitational interaction  energy for  the different polytropic  structures (the
green  and blue  crosses  in  the figure  are  indistinguishable).  This  simple
example demonstrates  that it is  mandatory to  account for the  contribution of
substructures  to  correctly  evaluate  interaction  energies  in  astrophysical
structures.

\section{The nonideal Friedmann equations}

As suggested in Tremblin {\it et al.}\cite{tremblin}, Einstein equations may not
be  suited for  the  study of  correlated  fluids for  two  reasons. First,  the
Newtonian  limit  gives   back  the  Poisson  equation  that   is  a  mean-field
approximation  ignoring  the  presence  of  correlations.   Second,  it  can  be
explicitly shown that  the equivalence principle for a  fluid implicitly assumes
the absence  of correlations.   We therefore  propose to  rely on  the Newtonian
demonstration of the Friedmann equations in order  to get an intuition of how to
design  a nonideal  version  that takes  into  account correlations.   Following
Peacock\cite{peacock},  we assume  an ensemble  of particles  of mass  $m$ in  a
volume $V$ of radius  $R$, whose fluid density is assumed  to be homogeneous and
isotropic at very large scales, but can be correlated at smaller scales, that is
$P_1(\vec{r})=1/V$              but              $P_2(\vec{r},\vec{r}^\prime)\ne
P_1(\vec{r})P_1(\vec{r}^\prime)$.  The  total mechanical  energy of  this system
under its N-body or fluid form is given by
\begin{eqnarray}\label{eq:emec}
  E_m = \langle\ m v^2/2\rangle + \langle H_\mathrm{int,Nbody} \rangle.
\end{eqnarray}
Similarly  to  $\alpha_{\mathrm{ih}}$  in Sect.~\ref{sect:polytrop},  we  define
$\alpha_{\mathrm{ni}}$ by
\begin{equation}\label{eq:alpha_ni_general}
  \alpha_{\mathrm{ni}}  =  \frac{\iint_{V,V}  \phi(|\vec{r}  -  \vec{r}^\prime|)
    P_2(\vec{r},  \vec{r}^\prime)  dV  dV^\prime} {\iint_{V,V}  \phi(|\vec{r}  -
    \vec{r}^\prime|) P_1(\vec{r}) P_1(\vec{r}^\prime) dV dV^\prime}.
\end{equation}
The ideal (uncorrelated) fluid hypothesis corresponds to $\alpha_{\mathrm{ni}} =
1$. Eq.~\ref{eq:emec} can then be rewritten as
\begin{eqnarray}
  E_m    &=&    \langle\    m   v^2/2\rangle    +\alpha_{\mathrm{ni}}    \langle
  H_\mathrm{int,ideal} \rangle.
\end{eqnarray}
Following   Peacock\cite{peacock},  we   take  $   \langle\  m   v^2/2\rangle  =
M\dot{R}^2/2$ and  $\langle H_\mathrm{int,ideal} \rangle =  -GM^2/R$ and rewrite
this equation as
\begin{eqnarray}
  E_m/M = \frac{\dot{R}^2}{2}- \frac{4\pi}{3}\alpha_{\mathrm{ni}}G \rho_m R^2.
\end{eqnarray}
Rearranging the different terms, we obtain
\begin{eqnarray}\label{eq:hubble_ni}
  \frac{\dot{R}^2}{R^2}      -\frac{2E_m/M}{R^2}      =     \frac{8\pi      G}{3
    c^2}\alpha_{\mathrm{ni}} \rho_m c^2 .
\end{eqnarray}
In the  uncorrelated case, $\alpha_{\mathrm{ni}}=1$, and  with the substitutions
$R\rightarrow a$  and $-E_m/M \rightarrow  K$, we recognize the  first Friedmann
equation.  Therefore, Eq.~\ref{eq:hubble_ni}  indicates that correlations should
be accounted for  through a multiplicative factor  $\alpha_{\mathrm{ni}}$ of the
energy density  for a proper  nonideal generalization of the  Friedmann equation
(this  could also  be  seen  as a  multiplicative  factor  of the  gravitational
constant). The last step in obtaining  a nonideal first Friedmann equation is to
take  the  thermodynamic  limit in  $\alpha_{\mathrm{ni}}$,  $(N,V)  \rightarrow
\infty$  keeping  $N/V=\rho$  constant.   This  can be  done  by  introducing  a
near-field approximation on a scale  $\lambda_{H}$, that is, by replacing $\phi$
by $\exp(-r/\lambda_H)\phi$  in Eq.~\ref{eq:alpha_ni_general},  or equivalently,
introducing  a cutoff  of  the  integrals at  a  radius $r_{\mathrm{bound}}$  of
approximately $\lambda_H$.  In the latter case, and assuming a large-scale flat,
homogeneous,  and isotropic  Universe, we  obtain the  following first  nonideal
Friedmann equation:
\begin{equation}
  H_\mathrm{ni}^2  =   \frac{8  \pi  G}{3c^2}\alpha_\mathrm{ni}   \rho_b,  \quad
  \mathrm{with} \quad \alpha_\mathrm{ni} = \frac{\int_0^{r_\mathrm{bound}} g(r)r
    dr }{\int_0^{r_\mathrm{bound}} r dr}.
\end{equation}
We can  now study the  acceleration of the expansion.   Because we can  link the
value of  the Hubble parameter  to the  nonideal amplification induced  by bound
substructures, $\alpha_\mathrm{ni}$, it is natural  to expect an acceleration of
the expansion linked to an increase in the densities of these structures, due to
the  ongoing  gravitational collapse.   In  order  to  get the  second  nonideal
Friedmann  equation, we  use the  first law  of thermodynamics  in an  expanding
universe with the derivation of the  first Friedmann equation.  The two nonideal
Friedmann equations (for $K=0$) in our formalism are given by
\begin{equation}
  H_\mathrm{ni}^2   =   \frac{8  \pi   G}{3c^2}\alpha_\mathrm{ni}   \rho_b,\quad
  \mathrm{and}                \quad               q_\mathrm{ni}                =
  -1-\frac{\dot{H}_\mathrm{ni}}{H_\mathrm{ni}^2}    \approx     \frac{1}{2}    -
  \frac{\dot{\alpha_\mathrm{ni}}}{2H_\mathrm{ni} \alpha_\mathrm{ni}}.
\end{equation}
It is  then natural to  obtain an acceleration of  the expansion induced  by the
formation of the large scale structures of the Universe which naturally produces
$\dot{\alpha}_\mathrm{ni}>0$  with an  increase  of the  absolute  value of  the
gravitational  interaction  energy   even  if  the  mean   density  $\rho_b$  is
decreasing.   Assuming the  nonideal  amplification factor  $\alpha_\mathrm{ni}$
entirely explains  the observed present-day  expansion, that is, it  varied from
$\alpha_\mathrm{ni}=1$ to  $\alpha_\mathrm{ni} = \rho_c/\rho_b \approx  20$ over
the age of  the universe, $t_u \sim$14 Gyr, we  can obtain an order-of-magnitude
analytical expression,
\begin{equation}\label{eq:qni_rough}
  q_\mathrm{ni} \approx \frac{1}{2} - \frac{\ln(\rho_c/\rho_b)}{2 t_u H},
\end{equation}
which gives a  deceleration parameter $q_\mathrm{ni} =-1.06$  that is compatible
with the value based on type Ia supernovae $q=-1.0\pm0.4$ \cite{riess}.


\section*{References}

\begin{thebibliography}{99}
\bibitem{yvon}  J.   Yvon  in   La  th{\'e}orie   statistique  des   fluides  et
  l'{\'e}quation d'{\'e}tat, ed.  Actualit{\'e}s Scientifiques et Industrielles:
  Th{\'e}ories M{\'e}caniques (Hermann \& cie, 1935).

\bibitem{kirkwood}J.      G.      Kirkwood       and      E.      M.      Boggs,
  \Journal{\jchem}{10}{394}{1942}.

\bibitem{peacock}  J.   A.  Peacock  in  Cosmological   Physics,  ed.  Cambridge
  Astrophysics (Cambridge University Press, 1999).

\bibitem{tremblin}  P.   Tremblin  {\it   et  al.},   \Journal{\AA}{659}{A108}{2022}.

  \bibitem{riess} A. G. Riess {\it et al.}, \Journal{\AJ}{116}{1009}{1998}.

\end{thebibliography}
\end{document}